\documentclass[preprintnumbers,aps,prd,reprint,groupedaddress,nofootinbib]{revtex4-1}%,nobibnotes
\pdfoutput=1

\usepackage{booktabs}
\usepackage{footnote}

\usepackage{lipsum}
\usepackage{amsmath}    %Allows \begin{equation}, \begin{align}, \underbrace{2+2=4}_{\text{obvious}}
\usepackage{amssymb}    %Allows use math symbols
\usepackage{color}         %Allows {\color{red} Hello World} or \colorbox{blue}{Hello World}
\usepackage{graphicx}     %Allows \includegraphics{figure.pdf}
\usepackage[T1]{fontenc}
\usepackage{times}         %Use Times New Roman font
\usepackage{mathrsfs} %Allows \mathscr{HELLO}
\usepackage{myterms}
\usepackage{bm}
\usepackage{hyperref} %Automatically links \label and \ref commands; Always load last
\usepackage{multirow}
\hypersetup{
    colorlinks=true,       % false: boxed links; true: colored links
    linkcolor=blue,        % color of internal links
    citecolor=blue,        % color of links to bibliography
    filecolor=magenta,     % color of file links
    urlcolor=blue          % color of external links
}
\usepackage[all]{hypcap} %Link navagates to top of figure instead of caption (below fig)

\usepackage[none]{hyphenat} % One word not splited to two lines

\usepackage{color,xcolor,fancybox,epsf,rotating,colordvi}

\begin{document}
\preprint{\parbox{1.6in}{\noindent WHU-HEP-PH-TEV008}}

\title{Higgs decay to light scalars in the semi-constrained NMSSM}

\author{Shiquan Ma}
\email[]{shiqma@whu.edu.cn}
\affiliation{Center for Theoretical Physics, School of Physics and Technology, Wuhan University, Wuhan 430072, China}

\author{Kun Wang}
\email[]{wk2016@whu.edu.cn}
\affiliation{Center for Theoretical Physics, School of Physics and Technology, Wuhan University, Wuhan 430072, China}

\author{Jingya Zhu}
\email[]{zhujy@whu.edu.cn}
\affiliation{Center for Theoretical Physics, School of Physics and Technology, Wuhan University, Wuhan 430072, China}

\date{\today}

\begin{abstract}
The next-to minimal supersymmetric standard model (NMSSM) with non-universal Higgs masses, or the semi-constrained NMSSM (scNMSSM), extend the minimal supersymmetric standard model (MSSM) by a singlet superfield and assume universal conditions except for the Higgs sector.
It can not only keep the simpleness and grace of the fully constrained MSSM and NMSSM, and relax the tension that they face after the 125-GeV Higgs boson discovered, but also predict an exotic phenomenon that Higgs decay to a pair of light singlet-dominated scalars ($10\!\sim\! 60\;{\rm GeV}$).
This condition can be classified to three scenarios according to the identities of the SM-like Higgs and the light scalar: (i) the light scalar is CP-odd, and the SM-like Higgs is $h_2$; (ii) the light scalar is CP-odd, and the SM-like Higgs is $h_1$; (iii) the light scalar is CP-even, and the SM-like Higgs is $h_2$.
In this work, we compare the three scenarios, checking the interesting parameter schemes that lead to the scenarios, the mixing levels of the doublets and singlets, the tri-scalar coupling between the SM-like Higgs and a pair of light scalars, the branching ratio of Higgs decay to the light scalars, and sensitivities in hunting for the exotic decay at the HL-LHC and the future lepton colliders such as CEPC, FCC-ee, and ILC.
% the detections at the hadron colliders and future lepton colliders.
% We find that: ($\romannumeral1$)around $420\fbm$ on HL-LHC is enough to observe this new scalar if it's light enough that the best channel is $Br(H_{SM}\to ss \to 2\mu\tau)$; ($\romannumeral2$)as for electronic colliders, $740fb^{-1}$ on FCC-ee, $170\fbm$ on CEPC, $100\fbm$ on ILC is enough. Both $Br(H_{SM}\to ss \to 4b)$ and $Br(H_{SM}\to ss \to 4\tau)$ channels are attractive.
\end{abstract}

\maketitle

%==================================
% Introduction
%==================================
\section{Introduction}
\label{sec:intro}

In 2012 a new boson of about $125\GeV$ was discovered at the LHC \cite{Aad:2012tfa, Chatrchyan:2012xdj}, and in later years it was verified as the SM-like Higgs boson with more and more data \cite{Khachatryan:2016vau, Sirunyan:2018koj, Aad:2019mbh, CMS-PAS-HIG-19-005, Sopczak:2020vrs}.
But some other questions still exist, e.g., whether another scalar survives in the low mass region, and whether there is exotic Higgs decay to light scalars.
Before the LHC, for the low integrated luminosity (IL) the LEP did not exclude a light scalar with a smaller production rate than the SM-like Higgs \cite{Barate:2003sz}.
The CMS(ATLAS) collaboration searched for resonances directly
% in di-photon channel in the $65(70)\sim110\GeV$ \cite{CMS:2017yta, ATLAS:2018xad},
in $bj\mu\mu$ channel in the $10\!\sim\!60$ ($20\!\sim\!70$) GeV \cite{Sirunyan:2018wim, ATLAS-CONF-2019-036}.
The two collaborations also searched for the exotic Higgs decay to light resonances in final states with $b\bar{b}\tau^+\tau^-$ \cite{Sirunyan:2018pzn}, $b\bar{b}\mu^+\mu^-$ \cite{ Aaboud:2018esj, Sirunyan:2018mot}, $\mu^+\mu^-\tau^+\tau^-$ \cite{Sirunyan:2020eum, Sirunyan:2018mbx, Sirunyan:2019gou}, $4\tau$ \cite{Khachatryan:2017mnf, Sirunyan:2019gou}, $4\mu$ \cite{CMS-PAS-HIG-16-035, CMS-PAS-HIG-18-003, Aaboud:2018fvk}, $4b$ \cite{Aaboud:2018iil}, $\gamma\gamma gg$ \cite{Aaboud:2018gmx}, $4\gamma$ \cite{ATLAS-CONF-2012-079}.
But there is still sufficient space left of physics on the exotic decay.
For example, in the $b\bar{b}\tau^+\tau^-$ channel reported by CMS collaboration \cite{Sirunyan:2018pzn}, the $95\%$ exclusion limit is $3\%$ at least in the $20\sim60\GeV$ region.
But according to simulations, the future limits can be $0.3\%$ at the High-Luminosity program of the Large Hadron Collider (HL-LHC) \cite{CMS-PAS-FTR-18-035}, $0.04\%$ at the Circular Electron Positron Collider (CEPC), and $0.02\%$ at the Future Circular Colliders in $e^+e^-$ collisions (FCC-ee) \cite{An:2018dwb, Liu:2016zki}.

This exotic Higgs decay to light scalars can be motivated in many theories beyond the Standard Model (BSM) \cite{Curtin:2013fra}, e.g., the next-to minimal supersymmetric standard model (NMSSM), the simplest little Higgs model, the minimal dilaton model, the two-Higgs-doublet model, the next-to two-Higgs-doublet model, the singlet extension of the SM, etc.
Several phenomenological studies on the exotic decay exist in these models \cite{Dermisek:2005gg, Dermisek:2006wr, Dermisek:2006py, Carena:2007jk, Cheung:2007sva, Cao:2013gba, Cheung:2007sva, Han:2013ic, Cao:2013cfa, LiuLiJia:2019kye, Han:2018bni, Chun:2017yob, Bernon:2014nxa, Engeln:2018mbg, Haisch:2018kqx, Liu:2016ahc}.

The NMSSM extend the MSSM by a singlet superfield $\hat{S}$, solving the $\mu$-problem of it, and relax its fine-tuning tension after Higgs discovered in 2012 \cite{King:2012tr, Benbrik:2012rm, Cao:2012fz, Cao:2012yn, Kang:2012sy, King:2012is, Ellwanger:2011aa}.
However, as supersymmetric (SUSY) models, the MSSM and NMSSM both suffer from a huge parameter space of over 100 dimensions.
In most studies, some parameters are assumed equal at low-energy scale manually, leaving only about 10 free ones, and without considering the  Renormalization Group Equations (RGEs) running from high scales \cite{King:2012tr, Benbrik:2012rm, Cao:2012fz, Cao:2012yn, Kang:2012sy, King:2012is, Ellwanger:2011aa}.
In Ref.\cite{Cao:2013gba} a Higgs boson of $125\GeV$ decay to light scalars were studied in the NMSSM with parameters set in this way.
While in constrained models, congeneric parameters are assumed universal at the Grand Unified Theoretical (GUT) scale, leaving only four free parameters in the fully-constrained MSSM (CMSSM) and four or five in the fully-constrained NMSSM (CNMSSM) \cite{Ellwanger:2010es, LopezFogliani:2009np, Belanger:2008nt, Djouadi:2008uj, Ellwanger:2008ya, Hugonie:2007vd, Kowalska:2012gs, Gunion:2012zd}.
However, it was found that CMSSM and CNMSSM were nearly excluded considering the $125\GeV$ Higgs data, high mass bounds of gluino and squarks in the first two generations, muon g-2, dark matter relic density and detections \cite{Gunion:2012zd, Kowalska:2012gs, Cao:2011sn, Ellis:2012aa, Bechtle:2015nua, Athron:2017qdc, Wang:2018vrr}.

The semi-constrained NMSSM (scNMSSM) relaxes the unified conditions of the Higgs sector at the GUT scale, thus it is also called NMSSM with non-universal Higgs mass (NUHM) \cite{Das:2013ta, Ellwanger:2014dfa, Wang:2018vxp, Nakamura:2015sya}.
It not only keeps the simpleness and grace of the CMSSM and CNMSSM, but also relax the tension that they facing after the SM-like Higgs discovered \cite{Wang:2020dtb}, and also predicts interesting light particles such as a singlino-like neutralino \cite{Wang:2020tap}, and light Higgsino-dominated NLSPs \cite{Wang:2019biy, Ellwanger:2018zxt, Ellwanger:2016sur}, etc.
In this work, we study the scenarios in the scNMSSM with a light scalar of $10\sim60\GeV$, and the detections of exotic Higgs decay to a pair of it.

The main point of this paper is listed as follows.
In \sref{sec:ana}, we introduce the model briefly and give some related analytic formulas.
In \sref{sec:num} we present in detail the numerical calculations and discussions.
Finally, we draw our conclusions in \sref{sec:con}.

\section{The model and analytic calculations}
\label{sec:ana}

The superpotential of NMSSM, with $\mathbb{Z}_3$ symmetry, is written as \cite{Maniatis:2009re}
\begin{equation}\label{F-term}
    W=W_{\rm Yuk}+\lambda \hat{S} \hat{H}_{u}\cdot\hat{H}_{d}+\frac{1}{3}\kappa \hat{S}^3\,,
\end{equation}
from which the so-called F-terms of the Higgs potential can be derived as
\begin{equation}
    V_{\rm F}=|\lambda S|^2(|H_u|^2+|H_d|^2)+|\lambda H_u\cdot H_d+\kappa S^2|^2 \,.
\end{equation}
The D-terms is the same as in the MSSM
\begin{equation}\label{D-term}
    V_{\rm D} =\frac{1}{8}\left(g_1^2+g_2^2\right)\left(|H_d|^2-|H_u|^2\right)^2 +\frac{1}{2}g_2^2\left|H^{\dagger}_u H_d\right|^2 \,,
\end{equation}
where $g_1$ and $g_2$ are the gauge couplings of $U(1)_Y$ and $SU(2)_L$ respectively.
Without considering the SUSY-breaking mechanism, at a low-energy scale the soft-breaking terms can be imposed manually to the Lagrangian.
In the Higgs sector these terms corresponding to the superpotential are
\begin{eqnarray}\label{soft-term}
   V_{\rm soft}&=&M^2_{H_u}|H_u|^2+M^2_{H_d}|H_d|^2+M^2_S|S|^2 \nonumber\\
   &&+\left(\lambda A_{\lambda}SH_u\cdot H_d+\frac{1}{3}\kappa A_{\kappa}S^3+h.c.\right) \,,
\end{eqnarray}
where $M^2_{H_u},\, M^2_{H_u},\, M^2_{S}$ are the soft masses of Higgs fields $H_u,\, H_d,\,S$, and $A_\lambda,\, A_\kappa$ are the trilinear couplings at $M_{\rm SUSY}$ scale respectively.
However, in the scNMSSM the SUSY breaking is mediated by gravity, thus the soft-parameters at $M_{\rm SUSY}$ scale are running naturally from the GUT scale complying with the RGEs.

At electroweak symmetry breaking, $H_u$, $H_d$ and $S$ get their vacuum expectation values (VEVs) $v_u$ , $v_d$ and $v_s$ respectively, with $\tan\beta\equiv v_u/v_d$, $\sqrt{v_u^2+v_d^2}\approx173\GeV$, and $\mu_{\rm eff}\equiv \lambda v_s$.
Then they can be written as
\begin{eqnarray}
&&H_u=\left(
      \begin{array}{c}
         H_u^+ \\
         v_u+\frac{\phi_1+i\varphi_1}{\sqrt{2}} \\
      \end{array}
    \right), \quad
H_d=\left(
      \begin{array}{c}
        v_d+\frac{\phi_2+i\varphi_2}{\sqrt{2}} \\
        H_d^- \\
      \end{array}
    \right), \quad \nonumber \\
&&~~ S=v_s+\frac{\phi_3+i\varphi_3}{\sqrt{2}}.
\end{eqnarray}
The Lagrangian is consist of the F-terms, D-terms, and soft-breaking terms, so with the above equations one can get the tree-level squared-mass matrix of CP-even Higgses in the base $\{\phi_1, \phi_2, \phi_3\}$ and CP-odd Higgses in the base $\{\varphi_1, \varphi_2, \varphi_3\}$ \cite{Maniatis:2009re}.
After diagonalizing the mass squared matrixes including loop corrections \cite{Carena:2015moc}, one can get the mass-eigenstate Higgses (three CP-even ones $h_{1,2,3}$ and two CP-odd ones $a_{1,2}$, in mass order) from the gauge-eigenstate ones ($\phi_{1,2,3}, \varphi_{1,2,3}$):
\begin{eqnarray}
&& \quad h_i=S_{ik}\, \phi_k, \quad a_j=P_{jk}\, \varphi_k \,,
\end{eqnarray}
where $S_{ik}, P_{jk}$ are the corresponding components of $\phi_k$ in $h_i$ and $\varphi_k$ in $a_j$ respectively, with $i,k=1,2,3$ and $j=1,2$.

In the scNMSSM, the SM-like Higgs (hereafter denoted as $h$ uniformly) can be CP-even $h_1$ or $h_2$, and the light scalar (hereafter denoted as $s$ uniformly) can be CP-odd $a_1$ or CP-even $h_1$.
Then the couplings between the SM-like Higgs and a pair of light scalars $C_{hss}$ can be written at tree level as \cite{Ellwanger:2004xm}
\begin{eqnarray}
C_{h_2h_1h_1}^{\rm tree}
    &\!=\!&\frac{\lambda^2}{\sqrt{2}}
    \big[v_u(\Pi^{122}_{211}+\Pi^{133}_{211})
  \\
    &&+v_d(\Pi^{211}_{211}+\Pi^{233}_{211})
    +v_s(\Pi^{311}_{211}+\Pi^{322}_{211})
    \big]
  \nonumber\\
    &&-\frac{\lambda\kappa}{\sqrt{2}}
    \bigl(v_u\Pi^{323}_{211}+v_d\Pi^{313}_{211}+2v_s\Pi^{123}_{211}\bigr)
  \nonumber\\
    &&+\sqrt{2}\kappa^2v_s \Pi^{333}_{211}-\frac{\lambda A_{\lambda}}{\sqrt{2}}\Pi^{123}_{211}+\frac{\kappa A_{\kappa}}{3\sqrt{2}}\Pi^{333}_{211}
  \nonumber\\
    &&+\frac{g^2}{2\sqrt{2}}
    \left[v_u (\Pi^{111}_{211}-\Pi^{122}_{211})-v_d (\Pi^{211}_{211}-\Pi^{222}_{211})
    \right] \,, \nonumber %\qquad
\end{eqnarray}
where
\begin{equation*}
\Pi^{ijk}_{211}=2S_{2i}S_{1j}S_{1k}+2S_{1i}S_{2j}S_{1k}+2S_{1i}S_{1j}S_{2k} \,;
\end{equation*}
or
\begin{eqnarray}
  C_{h_a a_1a_1}^{\rm tree}
    &=&\frac{\lambda^2}{\sqrt{2}}
    \big[v_u (\Pi^{122}_{a11}+\Pi^{133}_{a11})
   \\
    &&+v_d(\Pi^{211}_{a11}+\Pi^{233}_{a11})
    +v_s(\Pi^{311}_{a11}+\Pi^{322}_{a11})
    \big]
  \nonumber\\
    &&+\frac{\lambda\kappa}{\sqrt{2}}
    \big[v_u(\Pi^{233}_{a11}-2\Pi^{323}_{a11})+v_d (\Pi^{133}_{a11}-2\Pi^{313}_{a11})
  \nonumber\\
    &&+2v_s (\Pi^{312}_{a11}-\Pi^{123}_{a11}-\Pi^{213}_{a11})
    \big]
    +\sqrt{2}\kappa^2 v_s\Pi^{333}_{a11}
  \nonumber\\
    &&+\frac{\lambda A_{\lambda}}{\sqrt{2}}(\Pi^{123}_{a11}+\Pi^{213}_{a11}+\Pi^{312}_{a11})-\frac{\kappa A_{\kappa}}{3\sqrt{2}}\Pi^{333}_{a11}
  \nonumber\\
    &&+\frac{g^2}{2\sqrt{2}}
    \left[v_u(\Pi^{111}_{a11}-\Pi^{122}_{a11})-v_d (\Pi^{211}_{a11}-\Pi^{222}_{a11})
    \right] \,, \nonumber
\end{eqnarray}
where $\Pi^{ijk}_{a11}=2S_{ai}P_{1j}P_{1k}$ and $a=1,2$.
Thus the width of Higgs decay to a pair of light scalars can be given by
\begin{equation}
    \Gamma(h\to s s)=\frac{1}{32\pi m_{h}}C^2_{hss}\left({1-\frac{4m^2_{s}}{m^2_h}}\right)^{1/2} \,.
    \label{eq3}
\end{equation}

Then the light scalars continually decay to light SM particles, such as a pair of light quarks or leptons, or gluons or photons though loops.
The widths of light scalar decay to quarks and charged leptons at tree level are given by
\begin{eqnarray}
&&\Gamma(s\to l^+l^-) = \frac{\sqrt{2}G_F}{8\pi}m_s m^2_l \left({1-\frac{4m^2_l}{m^2_s}}\right)^{p/2} \,,
\label{eq1} \\
&&\Gamma(s\to q \bar{q}) = \frac{N_c G_F}{4\sqrt{2}\pi}C^2_{s q q}m_s m^2_q \left({1-\frac{4m^2_q}{m^2_s}}\right)^{p/2} \,,
\label{eq2}
\end{eqnarray}
where $p=1$ for CP-odd $s$, and $p=3$ for CP-even $s$.
And the couplings between light scalar and up-type or down-type quarks are given by
\begin{eqnarray}
C_{h_1t_L t^c_R}&=&\frac{m_t}{\sqrt{2}v \sin\beta}S_{11} \,,
\\
C_{h_1b_L b^c_R}&=&\frac{m_b}{\sqrt{2}v \cos\beta}S_{12} \,,
\\
C_{a_1t_L t^c_R}&=&i\frac{m_t}{\sqrt{2}v \sin\beta}P_{11} \,,
\\
C_{a_1b_L b^c_R}&=&i\frac{m_b}{\sqrt{2}v \cos\beta}P_{12} \,.
\end{eqnarray}

\section{Numerical calculations and discussions}
\label{sec:num}

In this work, we first scan the following parameter space with \textsc{NMSSMTools-5.5.2} \cite{Ellwanger:2004xm, Ellwanger:2005dv},
\begin{eqnarray}
&& 0\!<\!\lambda\!<\!0.7, \qquad 0\!<\!\kappa\!<\!0.7, \qquad 1\!<\!\tan\!\beta\!<\!30,
\nonumber \\
&& 100\!<\!\mu_{\rm eff}\!<\!200\GeV, \qquad 0\!<\!M_0\!<\!500\GeV,
\\
&& 0.5\!<\!M_{1/2}\!<\!2\TeV, \qquad |A_0|,\, |A_{\lambda}|,\, |A_{\kappa}|\!<\!10\TeV \,.
\qquad \nonumber
\end{eqnarray}

The constraints we imposed in our scan including:
(i) An SM-like Higgs of $123\!\!\sim\!\!127\GeV$, with signal strengths and couplings satisfying the current Higgs data \cite{Khachatryan:2016vau, Sirunyan:2018koj, Aad:2019mbh, CMS-PAS-HIG-19-005, Sopczak:2020vrs}.
(ii) Search results for exotic and invisible decay of the SM-like Higgs, and Higgs-like resonances in other mass regions, with \textsc{HiggsBounds-5.7.1} \cite{Bechtle:2008jh, Bechtle:2011sb, Bechtle:2013wla}.
(iii) The muon g-2 constraint, like in Ref.\cite{Wang:2020tap}.
(iv) The mass bounds of gluino and the first-two-generation squark over $2\TeV$, and search results for electroweakinos in multilepton channels \cite{Sirunyan:2018ubx}.
(vi) The dark matter relic density $\Omega h^2$ below $0.131$ \cite{Tanabashi:2018oca}, and the dark matter and nucleon scattering cross section below the upper limits in direct searches \cite{Aprile:2018dbl, Aprile:2019dbj}.
(vii) The theoretical constraints of vacuum stability and Landau pole.

After these constraints, the surviving samples can be categorized into three scenarios:
\begin{itemize}
    \item Scenario I: $h_2$ is the SM-like Higgs, and the light scalar $a_1$ is CP-odd;
    \item Scenario II: $h_1$ is the SM-like Higgs, and the light scalar $a_1$ is CP-odd;
    \item Scenario III: $h_2$ is the SM-like Higgs, and the light scalar $h_1$ is CP-even.
\end{itemize}
In Tab. \ref{tab1}, we list the ranges of parameters and light particle masses in the three scenarios.
From the table, one can see that the parameter ranges are nearly the same expect for $\lambda$, $\kappa$, and $A_\kappa$, but the mass spectrums for light particles are totally different.
\renewcommand{\arraystretch}{1.5}
\begin{table}[htbp]
  \centering
  \caption{The ranges of parameters and light particle masses in Scenario I, II and III.}
  \label{tab1}
    \begin{tabular}{|c|c|c|c|}
    \hline
 & Scenario I & Scenario II & Scenario III\\
 \hline
 $\lambda$ & $0\sim0.58$ & $0\sim 0.24$ & $0\sim 0.57$\\
 \hline
 $\kappa$ & $0\sim0.21$ & $0\sim0.67$ & $0\sim0.36$\\
 \hline
 $\tan\beta$ & $14\sim27$ & $10\sim28$ & $13\sim28$\\
 \hline
 $\mu_{\rm eff}\;[\rm GeV]$ & $103\sim200$ & $102\sim200$ & $102\sim200$\\
 \hline
  $M_0\;[\GeV]$ & $0\sim500$ & $0\sim500$ & $0\sim500$\\
 \hline
  $M_{1/2}\;[\TeV]$ & $1.06\sim1.47$ & $1.04\sim1.44$ & $1.05\sim1.47$\\
 \hline
  $A_0\;[\TeV]$ & $-2.8\sim0.2$ & $-3.2\sim-1.0$ & $-2.8\sim0.6$\\
 \hline
 $A_{\lambda} (M_{\rm GUT})\;[\TeV]$ & $1.3\sim9.4$ & $0.1\sim10$ & $1.1\sim9.8$\\
 \hline
 $A_{\kappa}(M_{\rm GUT})\;[\TeV]$ & $-0.02\sim5.4$ & $-0.02\sim0.9$ & $-0.7\sim5.7$\\
 \hline
 $A_{\lambda} (M_{\rm SUSY})\;[\rm TeV]$ & $2.0\sim10.1$ & $0.8\sim10.9$ & $1.6\sim10.2$\\
 \hline
 $A_{\kappa}(M_{\rm SUSY})\;[\rm GeV]$ & $-51\sim42$ & $-17\sim7$ & $-803\sim11$\\
 \hline
  $m_{\tilde{\chi}^0_1}\;[\GeV]$ & $3\sim129$ & $98\sim198$ & $3\sim190$\\
 \hline
  $m_{h_1}\;[\GeV]$ & $4\sim123$ & $123\sim127$ & $4\sim60$\\
 \hline
  $m_{h_2}\;[\GeV]$ & $123\sim127$ & $127\sim5058$ & $123\sim127$\\
 \hline
  $m_{a_1}\;[\GeV]$ & $4\sim60$ & $0.5\sim60$ & $3\sim697$\\
 \hline
    \end{tabular}
\end{table}

\begin{figure*}[!htbp]
\centering
\includegraphics[width=1.0\textwidth]{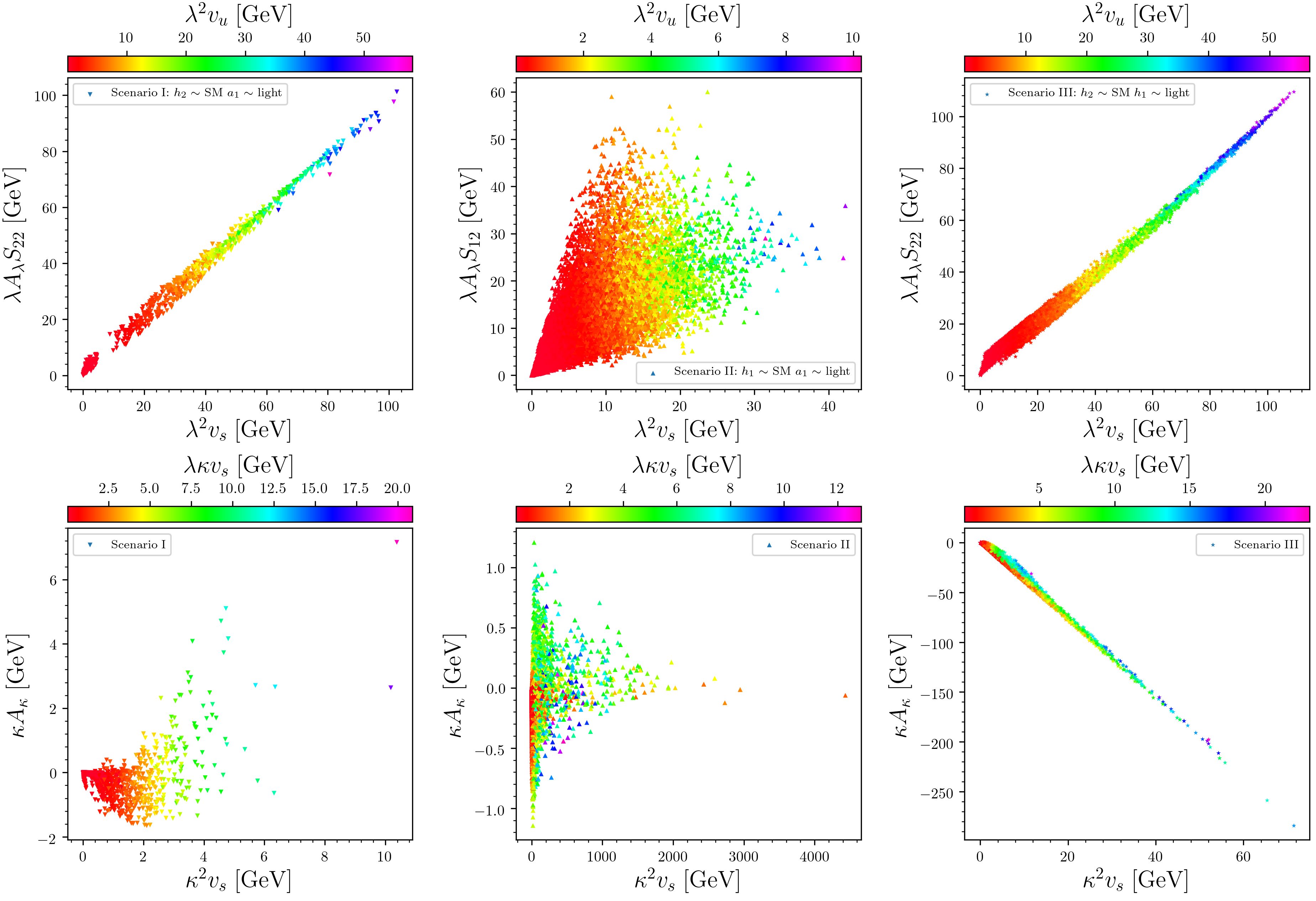}
\vspace{-0.4cm}
\caption{Surviving samples for the three scenarios in the $\lambda A_\lambda S_{i2}$ versus $\lambda^2 v_s$ (upper), where $S_{22}$ (left and right) and $S_{12}$ (middle) are the down-type-doublet component coefficient in the SM-like Higgs, and $\kappa A_\kappa$ versus $\kappa^2 v_s$ (lower) planes respectively.
Colors indicate $\lambda^2 v_u$ (upper) and $\lambda\kappa v_s$ (lower) respectively. }
\label{fig:1}
\end{figure*}

\begin{figure*}[!htbp]
\centering
\includegraphics[width=1.0\textwidth]{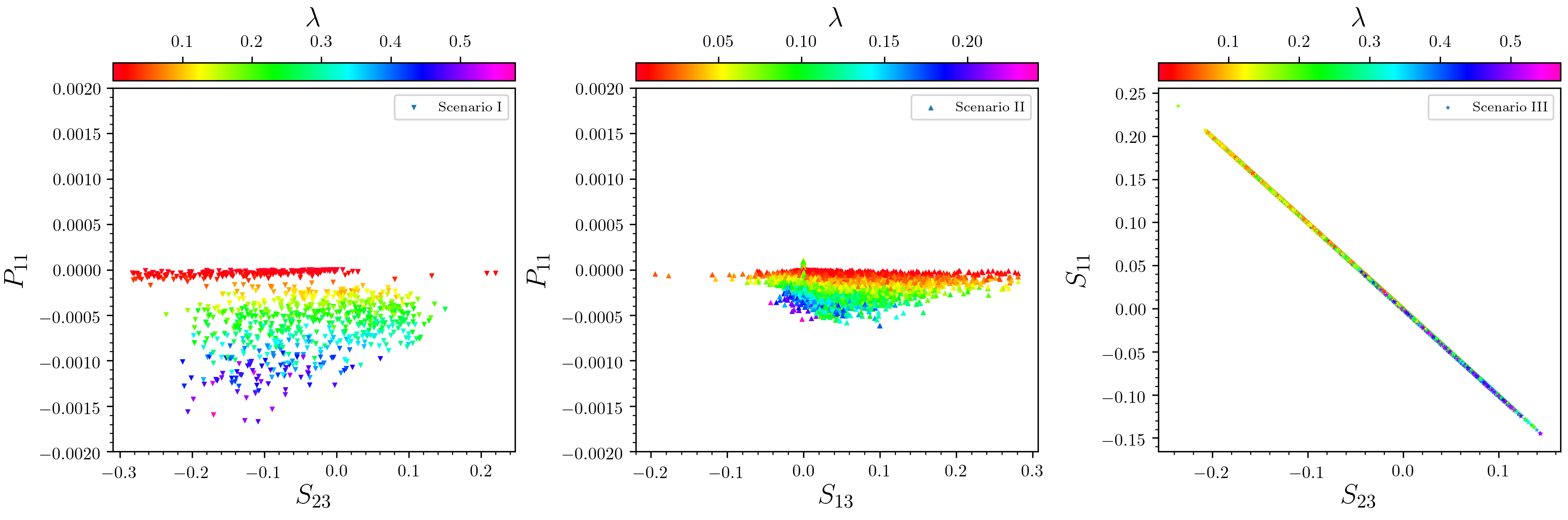}
\vspace{-0.4cm}
\caption{Surviving samples for the three scenarios in the $P_{11}$ versus $S_{23}$ (left), $P_{11}$ versus $S_{13}$ (middle), and $S_{11}$ versus $S_{23}$ (right) planes respectively,
where $S_{23}$ (left and right) and $S_{13}$ (middle) are the singlet component in the SM-like Higgs,
and $P_{11}$ (left and middle) and $S_{11}$ (right) are the up-type-doublet component of the light scalar respectively.
Colors indicate the parameter $\lambda$.}
\label{fig:2}
\end{figure*}

\begin{figure*}[!htbp]
\centering
\includegraphics[width=1.0\textwidth]{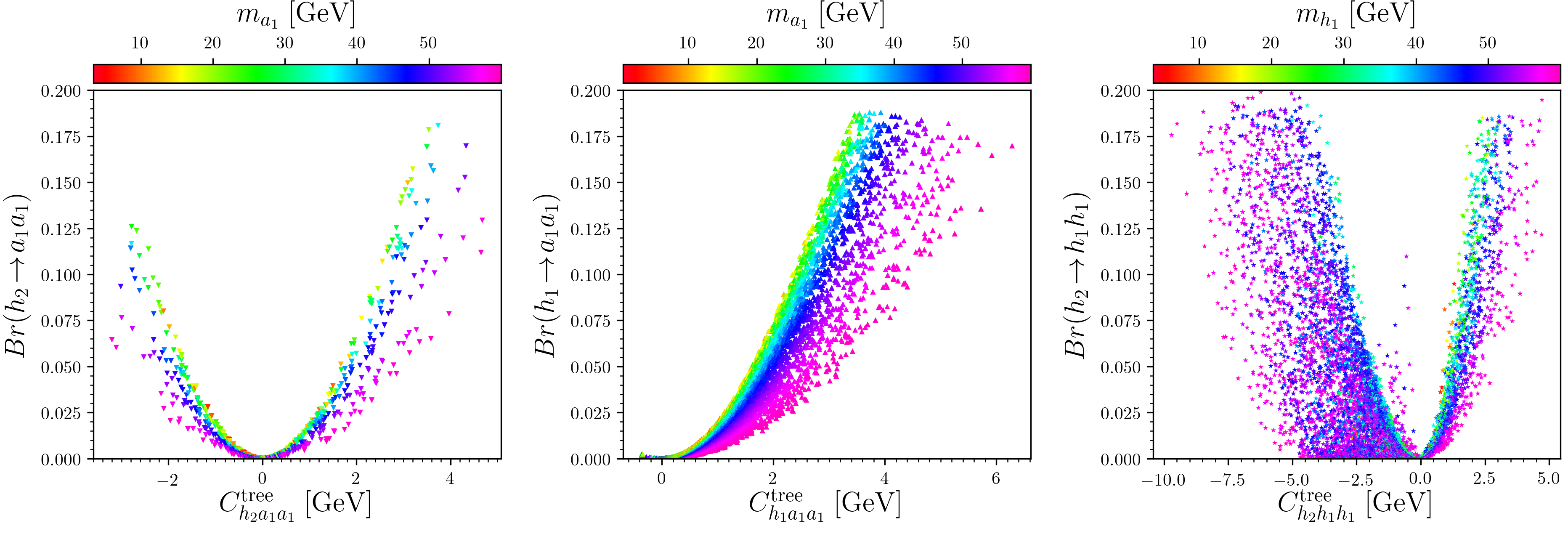}
\vspace{-0.4cm}
\caption{Surviving samples for the three scenarios in the exotic branching ratio $Br(h\!\to\!ss)$ versus the tri-scalar coupling $C_{hss}^{\rm tree}$ at tree level planes respectively, with colors indicate the mass of light Higgs $m_s$,
where $h$ denote the SM-like Higgs $h_2$ (left and right) and $h_1$ (middle),
and $s$ denote the light scalar $a_1$ (left and middle) and $h_1$ (right) respectively.}
\label{fig:3}
\end{figure*}

To study the different mechanisms of Higgs decay to light scalars in different scenarios, we recombine relevant parameters, and show them in Fig.\ref{fig:1}.
From this figure one can find that:
\begin{itemize}
  \item For Scenarios I and III, $\lambda A_{\lambda}S_{22} \!\approx\! \lambda^2v_s$, where $0.03\!\lesssim\! S_{22}\!\lesssim\!0.07$ is at the same order with $1/\tan\!\beta$, for the mass scale of the CP-odd doublet scalar $M_A \!\thicksim\! 2\mu_{\rm eff}/\sin\!2\beta \!\thicksim\! A_{\lambda} \!\gg\! \kappa v_s$ and $\tan\!\beta\!\gg\!1$ \cite{Cao:2013gba}.
      Thus the SM-like Higgs is up-type-doublet dominated.
  \item For Scenario I, $\kappa A_{\kappa}$, $k^2v_s$, and $\lambda\kappa v_s$ are at the same level of a few GeV; but for Scenario II, $\kappa^2 v_s$ can be as large as a few TeV for small $\lambda$ and large $\kappa$.
  \item Specially, for Scenario III, $\kappa A_{\kappa} \!\approx\! -4\kappa^2 v_s$, or $A_\kappa \!\approx\! -4\kappa v_s$.
\end{itemize}

According to the large data of the $125\GeV$ Higgs, and current null results searching for non-SM Higgs, the $125\GeV$ Higgs should be doublet dominated and the light scalar should be singlet dominated.
Therefore, both the singlet component in the SM-like Higgs and the doublet component in the light Higgs should be a small quantity generally.
We show how small they can be, and their relative scale in Fig.\ref{fig:2}.
From this figure, we can see as following for the three scenarios.
\begin{itemize}
    \item Scenario I: The up-type-doublet component of the light scalar $-\!0.0015 \!\lesssim\! P_{11} \!<\!0$ and is proportional to the parameter $\lambda$, thus the total doublet component of the light scalar $P_{1D}\!\equiv\! \sqrt{P_{11}^2+P_{12}^2}\!\thickapprox\! P_{11}\tan\beta \!\lesssim\!0.04$; while the singlet component of the SM-like Higgs $|S_{23}|\!\lesssim\!0.3$.
    \item Scenario II: The up-type-doublet component of the light scalar $-\!0.0006 \!\lesssim\!P_{11}\!<\!0 $ and is proportional to the parameter $\lambda$, thus total doublet component of the light scalar $0<P_{1D}\!\lesssim\!0.013$; while the singlet component in the SM-like Higgs $|S_{13}|\!\lesssim\!0.3$.
    \item Scenario III: The up-type-doublet component of the light scalar and the singlet component of the SM-like Higgs are anticorrelated $S_{11}\!\thickapprox\!-S_{23}$, and the range of them is $-0.15\!\lesssim\! S_{11}\!\lesssim\! 0.2$, with the sign related to the parameter $\lambda$. It also means that the mixing in the CP-even scalar sector is mainly between the singlet and the up-type doublet, and we checked that $0.03\!\lesssim\!S_{22}\!\lesssim0.07$ and $S_{12}\!\lesssim\!0.03$.
        Thus the SM-like Higgs is up-type doublet dominated, which is applicable in all three scenarios, with $S_{21}\!\approx\! 1$ in Scenario I and III and $S_{11}\!\approx\!1$ in Scenario II.
\end{itemize}
Considering the values of and correlations among parameters and component coefficients, the couplings between the SM-like Higgs and a pair of light scalars can be simplified as:
\begin{eqnarray}
C_{h_2a_1a_1}
&\simeq&
    \sqrt{2}\lambda^2v_u+\sqrt{2}\lambda A_{\lambda}P_{11}\tan\!\beta\,,
\label{ch2a1a1}
\\
C_{h_1a_1a_1}
&\simeq&
    \sqrt{2}\lambda^2v_u+\sqrt{2}\lambda A_{\lambda}P_{11}\tan\!\beta +2\sqrt{2}\kappa^2v_s S_{13} \,, \qquad
\label{ch1a1a1}
\\
C_{h_2h_1h_1}
&\simeq&
    \sqrt{2}\lambda^2v_u-\sqrt{2}\lambda A_{\lambda}S_{12} +\sqrt{2}\lambda^2v_s S_{11}
\nonumber \\
    &&+2\sqrt{2}\kappa^2 v_sS_{23} +\frac{3g^2}{\sqrt{2}}v_u S_{11}S_{11}
\nonumber \\
    &&-2\sqrt{2}\lambda\kappa v_s S_{12} \,.
\label{ch2h1h1}
\end{eqnarray}

In Fig.\ref{fig:3} we show the exotic branching ratio $Br(h\!\to\!ss)$ including one-loop correction correlated with the mass of the light scalar, and the coupling between the SM-like Higgs and a pair of the light scalars at tree level.
Since the 125 GeV Higgs is constrained to be very SM-like, its decay widths and branching ratios to SM particles cannot vary much.
Thus combined with Eq.(\ref{eq3}), it is natural that the branching ratios to light scalars are proportional to the square of the tri-scalar couplings.
The significant deviations for the negative-coupling samples in Scenario III are because of the one-loop correction of the stop loops,
\begin{eqnarray}
\Delta C_{h_2h_1h_1} &\simeq& S_{21} S_{11}^2 \frac{3\sqrt{2}m_t^4}{16\pi^2 v_u^3} \ln \left( \frac{m_{\tilde{t}_1}m_{\tilde{t}_2}}{m_t^2}\right),
\end{eqnarray}
which can be as large as $5\GeV$.
While for Scenario I and II, they are
\begin{eqnarray}
\Delta C_{h_2a_1a_1} &\simeq& S_{21} P_{11}^2 \frac{3\sqrt{2}m_t^4}{16\pi^2 v_u^3} \ln \left( \frac{m_{\tilde{t}_1}m_{\tilde{t}_2}}{m_t^2}\right),
\end{eqnarray}
\begin{eqnarray}
\Delta C_{h_1a_1a_1} &\simeq& S_{11} P_{11}^2 \frac{3\sqrt{2}m_t^4}{16\pi^2 v_u^3} \ln \left( \frac{m_{\tilde{t}_1}m_{\tilde{t}_2}}{m_t^2}\right).
\end{eqnarray}
% II (I), one can get that by only replace $S_{11}$ ($S_{21}S_{11}^2$) by $P_{11}$ ($S_{21}P_{11}^2$).
Since $P_{11}\!\ll\! S_{11}$ as seen form Fig.\ref{fig:2} the loop correction in Scenario I and II is much smaller than that in Scenario III.
In the following figures and discussions, we refer to the coupling $C_{hss}$ as including the one-loop correction $\Delta C_{hss}$ if without special instructions.

\subsection{Detections at the HL-LHC}

\begin{figure*}[!htbp]
\centering
\includegraphics[width=1.0\textwidth]{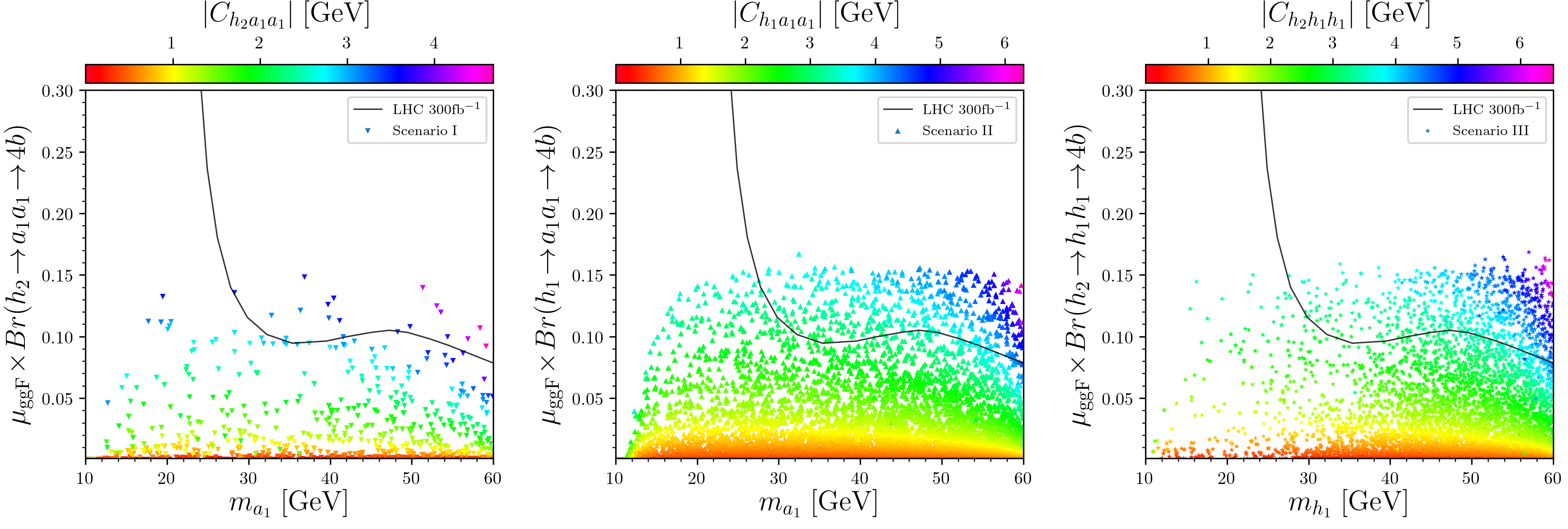}
\vspace{-0.4cm}
  \caption{Surviving samples for the three scenarios in the signal rate $\mu_{\rm Zh} \!\times\! Br(h\!\to\! ss\!\to\! 4b)$ versus the mass of light Higgs $m_s$ planes respectively, with colors indicate the tri-scalar coupling $C_{hss}$ including one-loop correction,
  where $h$ denote the SM-like Higgs $h_2$ (left and right) and $h_1$ (middle),
  and $s$ denote the light scalar $a_1$ (left and middle) and $h_1$ (right) respectively.
  The solid curves are the simulation result of the $95\%$ exclusion limit in the corresponding channel at the HL-LHC with $300\fbm$ \cite{Cao:2013gba}.
  }
\label{fig:4}
\end{figure*}

\begin{figure*}[!htbp]
\centering
\includegraphics[width=1.0\textwidth]{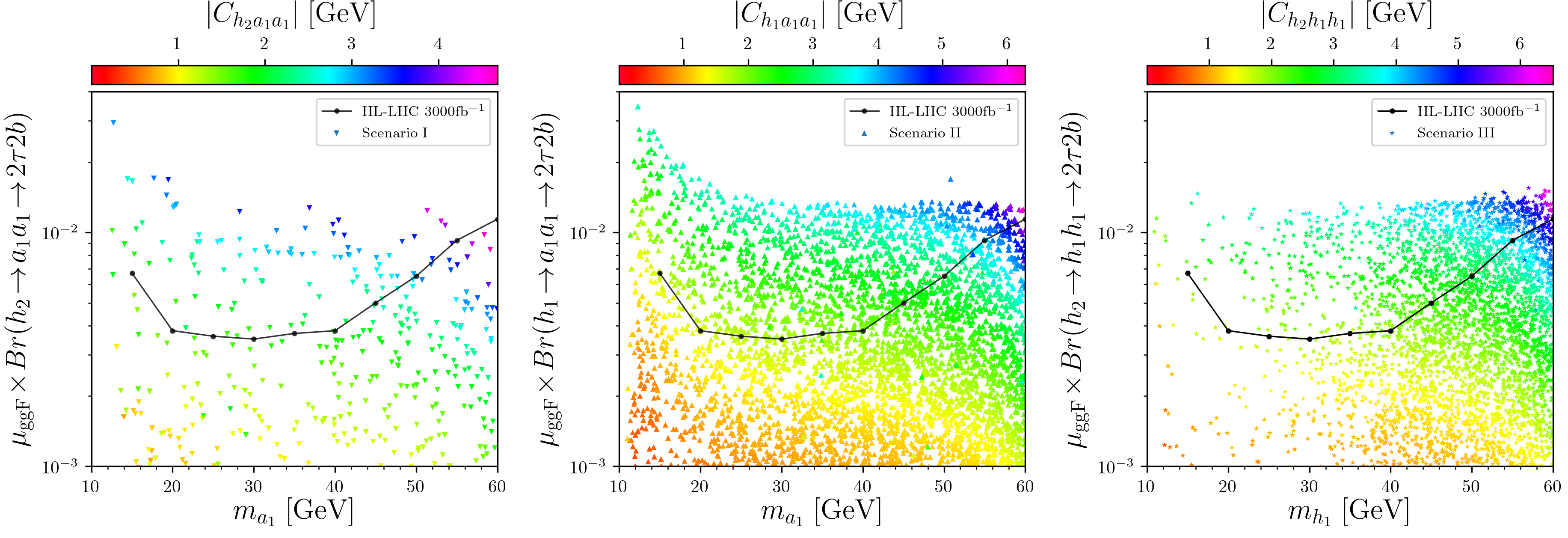}
\vspace{-0.4cm}
  \caption{Same as in Fig.\ref{fig:4}, but show the signal rate $\mu_{\rm ggF} \!\times\! Br(h\!\to\! ss\!\to\! 2\tau 2b)$, and $95\%$ exclusion limits in the corresponding channel at the HL-LHC with $3000\fbm$ \cite{CMS-PAS-FTR-18-035}.}
\label{fig:5}
\end{figure*}

\begin{figure*}[!htbp]
    \centering
    \includegraphics[width=1.0\textwidth]{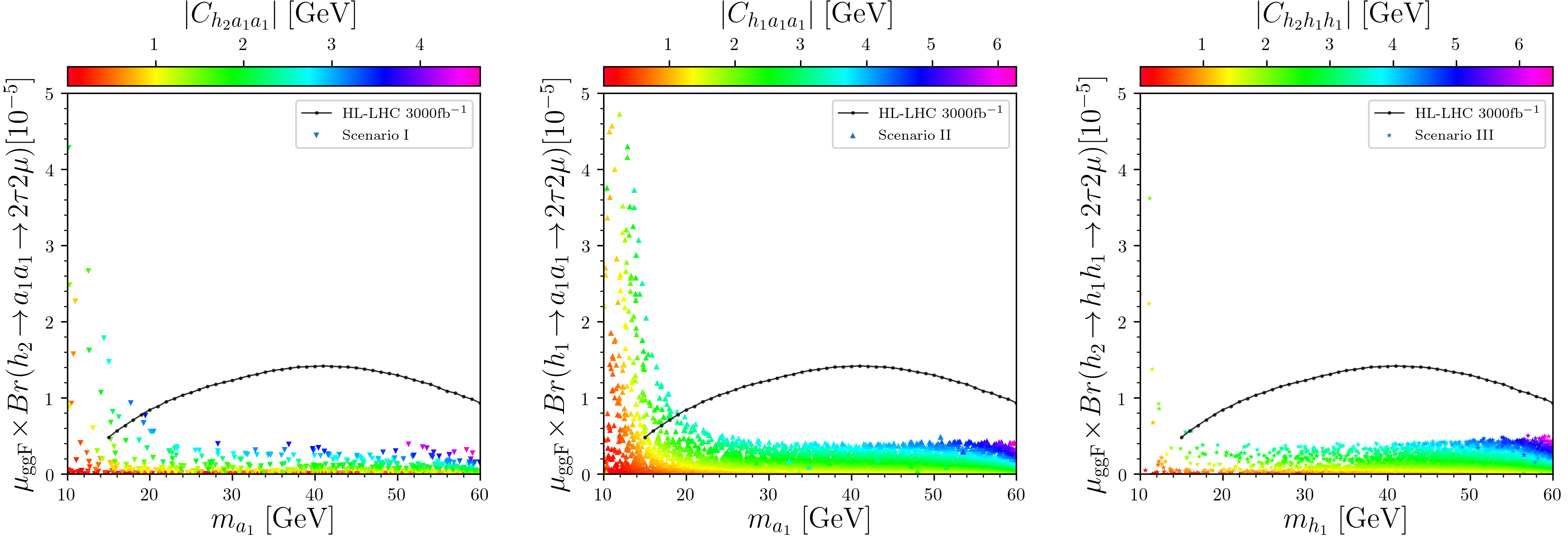}
    \vspace{-0.4cm}
  \caption{Same as in Fig.\ref{fig:4}, but show the signal rate $\mu_{\rm ggF} \!\times\! Br(h\!\to\! ss\!\to\! 2\tau 2\mu)$, and $95\%$ exclusion limits in the corresponding channel at the HL-LHC with $3000\fbm$ \cite{CMS-PAS-FTR-18-035}.}
    \label{fig:6}
\end{figure*}

At the LHC, the SM-like Higgs first can produce in gluon fusion (ggF), vector boson fusion (VBF), associated with vector boson ($\rm Wh$, $\rm Zh$), or associated with $t\bar{t}$ processes, where cross section in the ggF process is much larger than that of others.
Then the SM-like Higgs can decay to a pair of light scalars, and each scalar can then decay to a pair of fermions, or gluons, or photons.
The ATLAS and CMS collaborations have searched for these exotic decay mode in final states of $b\bar{b}\tau^+\tau^-$ \cite{Sirunyan:2018pzn}, $b\bar{b}\mu^+\mu^-$ \cite{ Aaboud:2018esj, Sirunyan:2018mot}, $\mu^+\mu^-\tau^+\tau^-$ \cite{Sirunyan:2020eum, Sirunyan:2018mbx, Sirunyan:2019gou}, $4\tau$ \cite{Khachatryan:2017mnf, Sirunyan:2019gou}, $4\mu$ \cite{CMS-PAS-HIG-16-035, CMS-PAS-HIG-18-003, Aaboud:2018fvk}, $4b$ \cite{Aaboud:2018iil}, $\gamma\gamma gg$ \cite{Aaboud:2018gmx}, $4\gamma$ \cite{ATLAS-CONF-2012-079}, etc.
These results are included in the constraints we considered.

As we checked, the main decay mode of the light scalar is usually to $b\bar{b}$ when $m_s\gtrsim 2m_b$.
However, the color backgrounds at the LHC are very large, thus minor $\rm Zh$ production process is used in detecting $h\!\!\to\!\! 2s \!\!\to\!\! 4b$, as well VBF used for $h\!\!\to\!\! 2s \!\!\to\!\! \gamma\gamma gg$.
For the other decay mode, the main production processes ggF can be used.
Considering the cross sections of production and branching ratios of decay, and the precisions of detection, we found the detections in $4b$, $2b2\tau$, and $2\tau 2\mu$ channels are important for the scNMSSM.
And the signal rates are $\mu_{\rm Zh} \!\times\! Br(h\!\to\! ss \!\to\! 4b)$, $\mu_{\rm ggF} \!\times\! Br(h\!\to\! ss \!\to\! 2b2\tau)$, and $\mu_{\rm ggF} \!\times\! Br(h\!\to\! ss \!\to\! 2\tau2\mu)$ respectively, where $\mu_{\rm ggF}$ and $\mu_{\rm Zh}$ are the ggF and $\rm Zh$ production rate normalized to their SM value respectively.

For detections of the exotic decay at the HL-LHC, we use the simulation results of $95\%$ exclusion limit in Refs.\cite{Cao:2013gba, CMS-PAS-FTR-18-035}.
Suppose with integrated luminosity of $L_0$, the $95\%$ exclusion limit for branching ratio in some channel is $Br_0$ in the simulation result, then for a sample in the model if the signal rate is $\mu_i\!\times\! Br$ ($i$ denote the production channel), the signal significance with integrated luminosity of $L$ will be
\begin{eqnarray}
ss = 2 \;\frac{\mu_i\!\times\! Br}{Br_0} \sqrt{\frac{L}{L_0}},
\end{eqnarray}
and the integrated luminosity needed to exclude the sample in the channel at $95\%$ confidence level (with $ss=2$) will be
\begin{eqnarray}
L_{\rm e}=L_0 \left(\frac{Br_0}{\mu_i\!\times\! Br}\right)^2,
\end{eqnarray}
and the integrated luminosity needed to discover the sample in the channel (with $ss=5$) will be
\begin{eqnarray}
L_{\rm d}= L_0 \left(\frac{5}{2}\right)^2 \left(\frac{Br_0}{\mu_i\!\times\! Br}\right)^2.
\end{eqnarray}

In Fig.\ref{fig:4}, \ref{fig:5}, and \ref{fig:6}, we show the signal rates for surviving samples in the three scenarios, and the $95\%$ exclusion limits \cite{Cao:2013gba, CMS-PAS-FTR-18-035} in the $4b$, $2b2\tau$, and $2\tau2\mu$ channels respectively.
From these figures one can see that
\begin{itemize}
  \item With the light scalar heavier than $30\GeV$, the easiest way to discover the exotic decay is in the $4b$ channel, and the minimal integrated luminosity needed to discover the decay in this channel can be $650\fbm$ for Scenario II.
  \item With the light scalar lighter than $20\GeV$, the $2\tau2\mu$ channel can be important, especially for samples in the Scenario II, and the minimal integrated luminosity needed to discover the decay in this channel can be $1000\fbm$.
  \item With the light scalar heavier than $2m_b$, chance all exist to discover the decay in the $2b2\tau$ channel, and the minimal integrated luminosity needed to discover the decay in this channel can be $1500\fbm$ for Scenario II.
\end{itemize}

\subsection{Detections at the future lepton colliders}

\begin{figure*}[!htbp]
    \centering
    \includegraphics[width=1.0\textwidth]{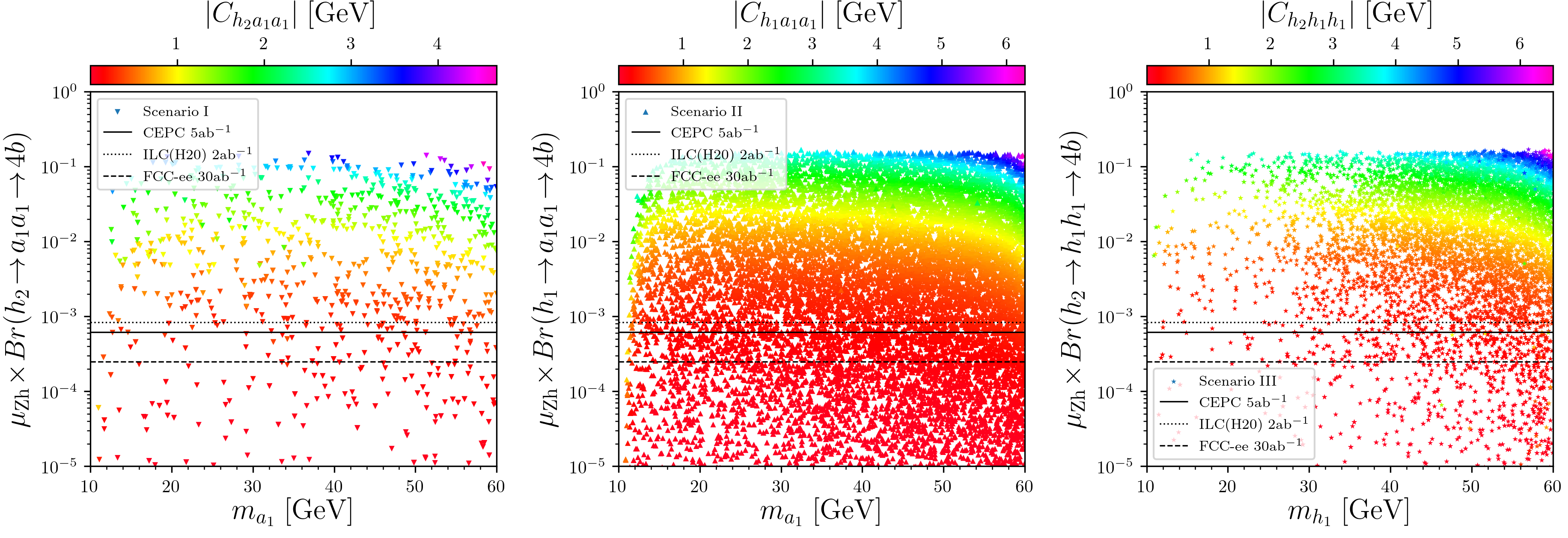}
    \vspace{-0.4cm}
\caption{Surviving samples for the three scenarios in the signal rate $\mu_{\rm Zh} \!\times\! Br(h\!\to\! ss\!\to\! 4b)$ versus the mass of light Higgs $m_s$ planes respectively, with colors indicate the tri-scalar coupling $C_{hss}$  including one-loop correction,
where $h$ denote the SM-like Higgs $h_2$ (left and right) and $h_1$ (middle),
and $s$ denote the light scalar $a_1$ (left and middle) and $h_1$ (right) respectively.
The solid, dashed, and dotted lines are the simulating result of $95\%$ exclusion limit in the corresponding channel at the CEPC with $5\,\rm{ab}^{-1}$, FCC-ee with $30\,\rm{ab}^{-1}$, and ILC with $2\,\rm{ab}^{-1}$ respectively \cite{Liu:2016zki}.
}
    \label{fig:7}
\end{figure*}

\begin{figure*}[!htbp]
    \centering
    \includegraphics[width=1.0\textwidth]{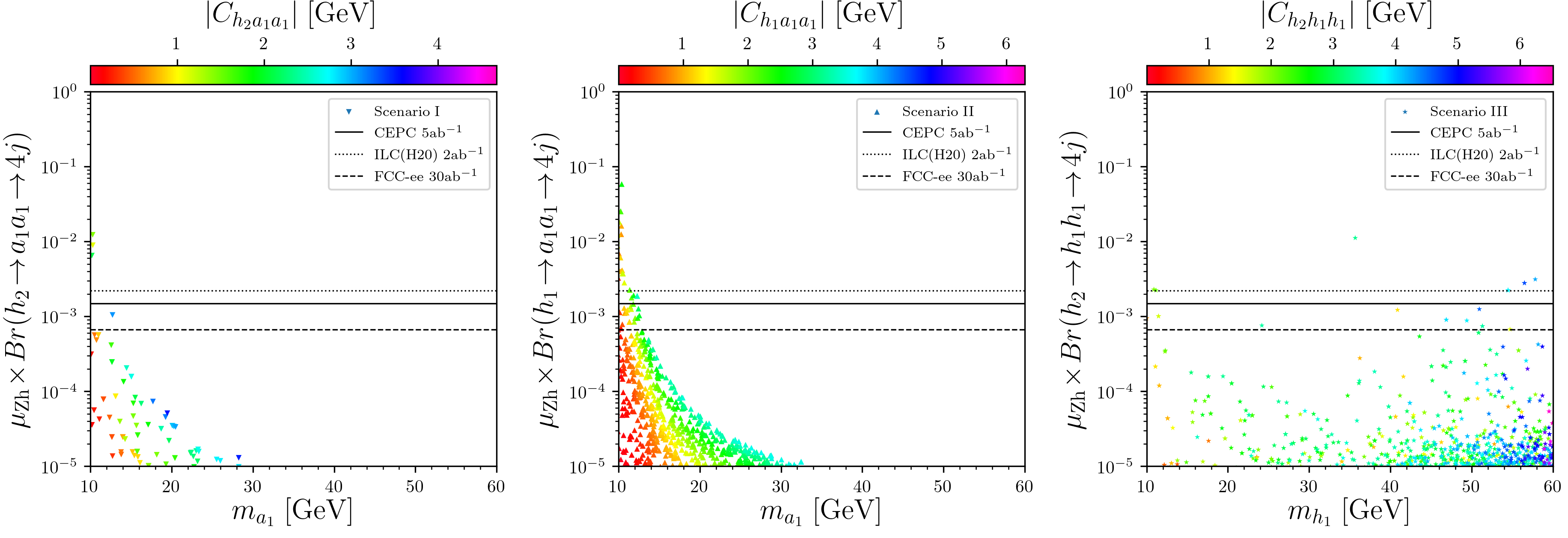}
    \vspace{-0.4cm}
\caption{Same as in Fig.\ref{fig:7}, but show the signal rate $\mu_{\rm Zh} \!\times\! Br(h\!\to\! ss\!\to\! 4j)$, and $95\%$ exclusion limits in the corresponding channel \cite{Liu:2016zki}.
The ``$4j$'' denotes four jets, including gluon and light quarks except for $b$.
}
\label{fig:8}
\end{figure*}

\begin{figure*}[!htbp]
    \centering
    \includegraphics[width=1.0\textwidth]{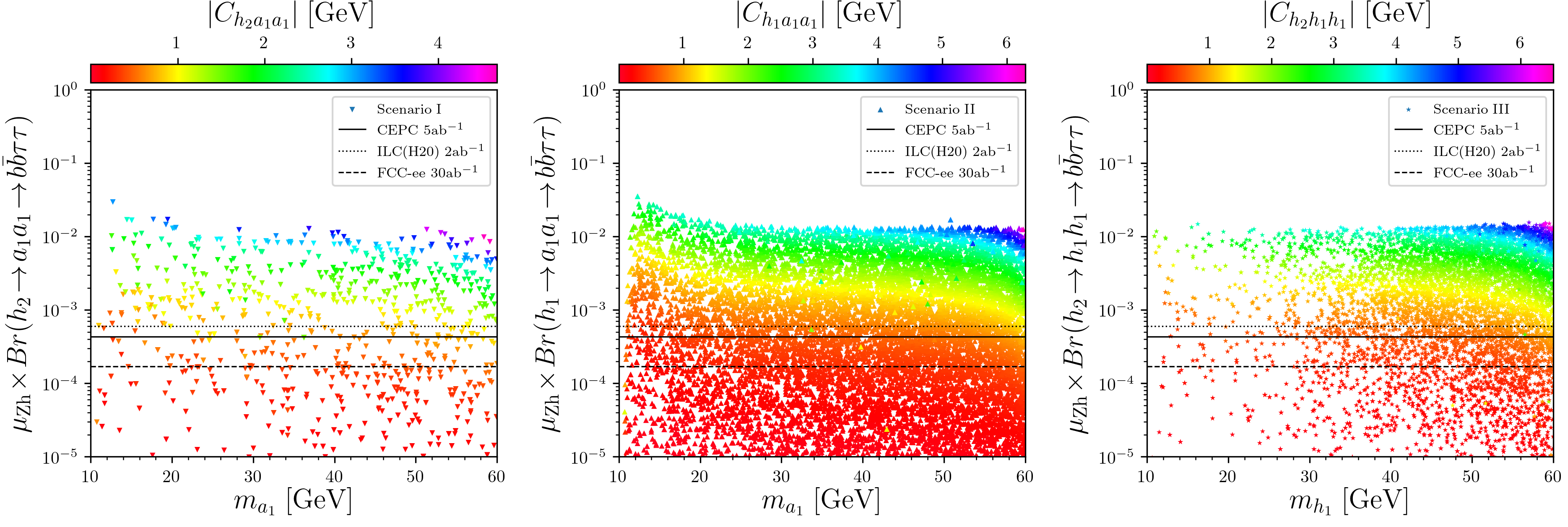}
    \vspace{-0.4cm}
\caption{Same as in Fig.\ref{fig:7}, but show the signal rate $\mu_{\rm Zh} \!\times\! Br(h\!\to\! ss\!\to\! 2b 2\tau)$, and $95\%$ exclusion limits in the corresponding channel \cite{Liu:2016zki}.
}
    \label{fig:9}
\end{figure*}

\begin{figure*}[!htbp]
    \centering
    \includegraphics[width=1.0\textwidth]{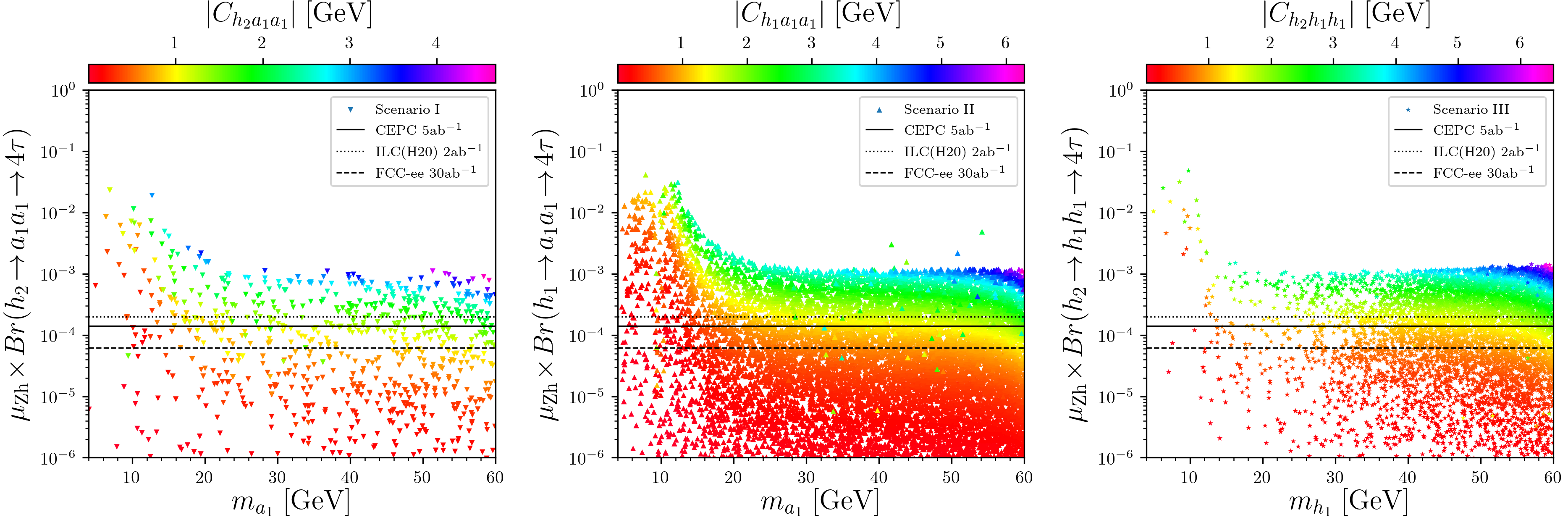}
    \vspace{-0.4cm}
\caption{
Same as in Fig.\ref{fig:7}, but show the signal rate $\mu_{\rm Zh} \!\times\! Br(h\!\to\! ss\!\to\! 4\tau)$, and $95\%$ exclusion limits in the corresponding channel \cite{Liu:2016zki}.
}
    \label{fig:10}
\end{figure*}

In future lepton colliders such as CEPC, FCC-ee, and International Linear Collider (ILC), the main production process of the SM-like Higgs is $\rm Zh$, and the color backgrounds are very little, thus these lepton colliders are powerful in detecting the exotic decay.
There have been simulation results in many channels, such as $4b$, $4j$, $2b2\tau$, $4\tau$, etc. \cite{Liu:2016zki}.
With the same method as in the last subsection, one can do similar analyses.

In Fig.\ref{fig:7}, \ref{fig:8}, \ref{fig:9}, and \ref{fig:10}, we show the signal rates for surviving samples in the three scenarios, and the $95\%$ exclusion limits at the CEPC, FCC-ee, and ILC, and in the $4b$, $4j$, $2b2\tau$, and $4\tau$ channels respectively \cite{Liu:2016zki}.
From these figures one can see that:
\begin{itemize}
  \item As in Fig.\ref{fig:7}, when the light scalar is heavier than about $15\GeV$ and the tri-scalar coupling is large enough, the branching ratio of $4b$ channel is significant.
      %where for Scenario I $C_{h_2a_1a_1}\!\gtrsim\!1.7\GeV$, for Scenario II $C_{h_1a_1a_1}\!\gtrsim\!2\GeV$, and for Scenario III $C_{h_2h_1h_1}\!\gtrsim\!5\GeV$.
      The minimal integrated luminosity needed to discover the decay in this channel can be $0.31\fbm$ for Scenario II and III at the ILC.
  \item As in Fig.\ref{fig:8}, for Scenario I and II, the exotic Higgs decay can be expected to be observed in the $4j$ channel when its mass is lighter than $11\GeV$. While for Scenario III, the light scalar available by CEPC can be as heavy as $40\GeV$.
      And the minimal integrated luminosity needed to discover the exotic decay in this channel can be $18\fbm$ for Scenario II at the ILC.
  \item As in Fig.\ref{fig:9} and \ref{fig:10}, the signal rates in $2b2\tau$ and $4\tau$ channel are in similar trends.
      The branching ratios are tiny before the light scalar reaches the mass threshold, the maximum of branching ratios occur around $m_s=12\GeV$, and the minimal integrated luminosity needed to discover the decay in $2b2\tau$ channel can be $3.6\fbm$ for Scenario II at the ILC, in $4\tau$ channel can be $0.22\fbm$ for Scenario III at the ILC.
\end{itemize}

\section{Conclusions}
\label{sec:con}

\renewcommand{\arraystretch}{1.5}
\begin{table*}[!htb]
  \centering
  \caption{The minimum integrated luminosity for discovering the exotic Higgs decay at the future colliders, where the ``@I, II, III'' means the three different scenarios.}
  \label{tab2}
  \setlength{\tabcolsep}{7mm}
    \begin{tabular}{ccccc}
    \hline
    \multirow{2}*{Deacy Mode}&
    \multicolumn{4}{c}{Futrue colliders}\cr
    \cline{2-5}
    &HL-LHC&CEPC&FCC-$ee$&ILC\cr
    \hline
\hline
($b\bar{b}$)($b\bar{b}$)&$650\fbm$(@II)&$0.42\fbm$(@III)&$0.41\fbm$(@III)&$0.31\fbm$(@II)\cr
($jj$)($jj$)&-&$21\fbm$(@II)&$18\fbm$(@II)&$25\fbm$(@II)\cr
($\tau^+\tau^-$)($\tau^+\tau^-$)&-&$0.26\fbm$(@III)&$0.22\fbm$(@III)&$0.31\fbm$(@III)\cr
($b\bar{b}$)($\tau^+\tau^-$)&$1500\fbm$(@II)&$4.6\fbm$(@II)&$3.6\fbm$(@II)&$4.4\fbm$(@II)\cr
($\mu^+\mu^-$)($\tau^+\tau^-$)&$1000\fbm$(@II)&-&-&-\cr
\hline
    \end{tabular}
\end{table*}

In this work, we have discussed the exotic Higgs decay to a pair of light scalars in the scNMSSM, or the NMSSM with NUHM.
First, we did a general scan over the nine-dimension parameter space of the scNMSSM, considering the theoretical constraints of vacuum stability and Landau pole, and experimental constraints of Higgs data, non-SM Higgs searches, muon g-2, sparticle searches, relic density and direct searches for dark matter, etc.
Then we found three scenarios with a light scalar of $10\!\sim\!60\GeV$:
(i) the light scalar is CP-odd, and the SM-like Higgs is $h_2$; (ii) the light scalar is CP-odd, and the SM-like Higgs is $h_1$; (iii) the light scalar is CP-even, and the SM-like Higgs is $h_2$.
For the three scenarios, we check the parameter schemes that lead to the scenarios, the mixing levels of the doublets and singlets, the tri-scalar coupling between the SM-like Higgs and a pair of light scalars, the branching ratio of Higgs decay to the light scalars, and the detections at the hadron colliders and future lepton colliders.

In this work, we compare the three scenarios, checking the interesting parameter schemes that lead to the scenarios, the mixing levels of the doublets and singlets, the tri-scalar coupling between the SM-like Higgs and a pair of light scalars, the branching ratio of Higgs decay to the light scalars, and the detections at the hadron colliders and future lepton colliders.

Finally, we draw following conclusions regarding a light scalar, and the exotic Higgs decay to a pair of it in the scNMSSM:
\begin{itemize}
  \item There are interesting different mechanisms in the three scenarios to tune parameters to get the small tri-scalar couplings.
  \item The singlet component of the SM-like Higgs in the three scenarios are at the same level of $\lesssim0.3$, and is roughly one-order larger than the doublet component of the light scalar in Scenario I and II.
  \item The coupling between the SM-like Higgs and a pair of light scalars at tree level is $-3\!\sim\!5$, $-1\!\sim\!6$ and $-10\!\sim\!5$ GeV for Scenario I, II, and III respectively.
  \item The stop-loop correction to the tri-scalar coupling in Scenario III can be a few GeV, much larger than that in Scenario I and II.
  \item The most effective way to discover the exotic decay at the future lepton collider is in the $4\tau$ channel; while that at the HL-LHC is $4b$ for the light scalar heavier than 30 GeV, or $2b2\tau$ and $2\tau2\mu$ for a lighter scalar.
\end{itemize}

In details, the minimal integrated luminosity needed to discover the exotic Higgs decay at the HL-LHC, CEPC, FCC-ee, and ILC are summarized in Tab.\ref{tab2}, and the tuning mechanisms in the three scenarios to get the small tri-scalar coupling can be seen from Figs. \ref{fig:1}, \ref{fig:2} and Eqs. (\ref{ch2a1a1}), (\ref{ch1a1a1}), (\ref{ch2h1h1}).

%----------------------------------------------------------------
% Acknowledgements
%----------------------------------------------------------------
\begin{acknowledgments}
\paragraph*{Acknowledgements.}
% We thank for discussions of, is grateful to for discussions of.
This work was supported by the National Natural Science Foundation of China (NNSFC) under grant No. 11605123.
\end{acknowledgments}

%----------------------------------------------------------------
% References
%----------------------------------------------------------------
\bibliography{ref}

\end{document}